\documentclass[aps,prl,reprint,twocolumn]{revtex4-2}
\usepackage{bbold}
\usepackage{amsmath,amsfonts,amsmath,mathtools,bbm,bm, amssymb}
\usepackage{graphicx}
\usepackage{comment}
\usepackage{tikz}
\usepackage{ragged2e}
\usepackage{caption}
\DeclareCaptionJustification{justified}{\justifying}
\usepackage{subcaption, stackengine}

\usepackage{braket}
\usepackage[colorlinks,linkcolor=blue,urlcolor=blue,citecolor=blue]{hyperref}
\usepackage[all]{hypcap}

\usepackage{lipsum}
\usepackage{blindtext}
\usepackage{enumitem}
\usepackage{xcolor}
\usepackage[normalem]{ulem}
\usepackage{orcidlink}

\usepackage{float}
\usepackage{array}
\setcitestyle{numbers,square}

\graphicspath{{Figure/PNG/}{Figure/PDF/}{Figure/EPS/}{Figure/TEX/}{Figure/}}

\newcommand\mydots{\hbox to 1em{.\hss.\hss.}}

\usepackage{titlesec}
\titleformat{\section}[runin]
  {\it}{}{1em}{\phantomsection}[]
\titleformat{\subsection}[runin]
  {\underline}{}{1em}{\phantomsection}[---]

\begin{document}

\title{Assembling extensive quantum Fisher information
in stabilizer systems}

\author{Arnau Lira-Solanilla~\orcidlink{0009-0000-1681-9334}}
\email{lirasola@thp.uni-koeln.de}
\affiliation{Institut f\"ur Theoretische Physik, Universit\"at zu K\"oln, Z\"ulpicher Stra{\ss}e 77, 50937 Cologne, Germany}

\author{Sreemayee Aditya~\orcidlink{}}
\affiliation{Institut f\"ur Theoretische Physik, Universit\"at zu K\"oln, Z\"ulpicher Stra{\ss}e 77, 50937 Cologne, Germany}

\author{Xhek Turkeshi~\orcidlink{0000-0003-1093-3771}}
\affiliation{Institut f\"ur Theoretische Physik, Universit\"at zu K\"oln, Z\"ulpicher Stra{\ss}e 77, 50937 Cologne, Germany}

\author{Silvia Pappalardi~\orcidlink{0000-0001-6931-8736}}
\email{pappalardi@thp.uni-koeln.de}
\affiliation{Institut f\"ur Theoretische Physik, Universit\"at zu K\"oln, Z\"ulpicher Stra{\ss}e 77, 50937 Cologne, Germany}

\date{\today}

\begin{abstract}
We introduce a systematic framework to construct nonlocal observables with extensive quantum Fisher information (QFI) density in stabilizer codes. The construction maps stabilizer generators to dual Ising spins whose correlators equal string order parameters, converting hidden nonlocal order into a metrologically accessible observable.
Applying this to monitored cluster codes and the toric code, we identify transitions in the QFI scaling from an extensive regime---where long-range string order prevails---to an intensive one driven by competing single-site measurements. 
\end{abstract}

\maketitle

\section{Introduction.}

The stabilizer formalism provides a compact algebraic description of large classes of quantum states and plays a central role in fault-tolerant quantum computation \cite{gottesman1997stabilizer, nielsen_chuang_2010, preskill2018quantumcomputingin}. 
In this approach, many-body quantum states are characterized by the set of mutually commuting operators (denoted stabilizers) that they simultaneously satisfy, naturally embedding global and topological information into the logical degrees of freedom. 
Because stabilizer codes encompass both symmetry-protected topological (SPT) states, such as cluster states \cite{raussendorf2001oneway}, and intrinsically topologically ordered phases, such as the toric code \cite{kitaev2003fault}, they bridge the encoding structure of quantum computation and the entanglement structure of many-body systems, attracting the interest of both fields.

A particularly rich setting within this framework is given by monitored dynamics, where projective measurements and unitary operations compete in quantum circuit settings. The resulting non-equilibrium phases are separated by sharp transitions in their entanglement structure: the celebrated measurement-induced phase transitions (MIPTs) \cite{skinner2019measurement, li2018quantum, li2019measurement, fisher2023random, potter2022entanglement, turkeshi2024error, bao2020theory, jian2020measurement, gullans2020dynamical, sierant2022measurement, noel2022measurement, koh2023measurement, hoke2023measurement}. These transitions establish entanglement itself as a meaningful order parameter and have motivated intense interest in the emergent statistical mechanics of monitored quantum matter. 
However, existing results predominantly concern bipartite entanglement, leaving open whether such systems host genuine multipartite entanglement and long-range order~\cite{amico2008entanglement, horodecki2009quantum, laflorencie2016quantum}.

\begin{figure}[t]
    \centering
    \includegraphics[width = \linewidth]{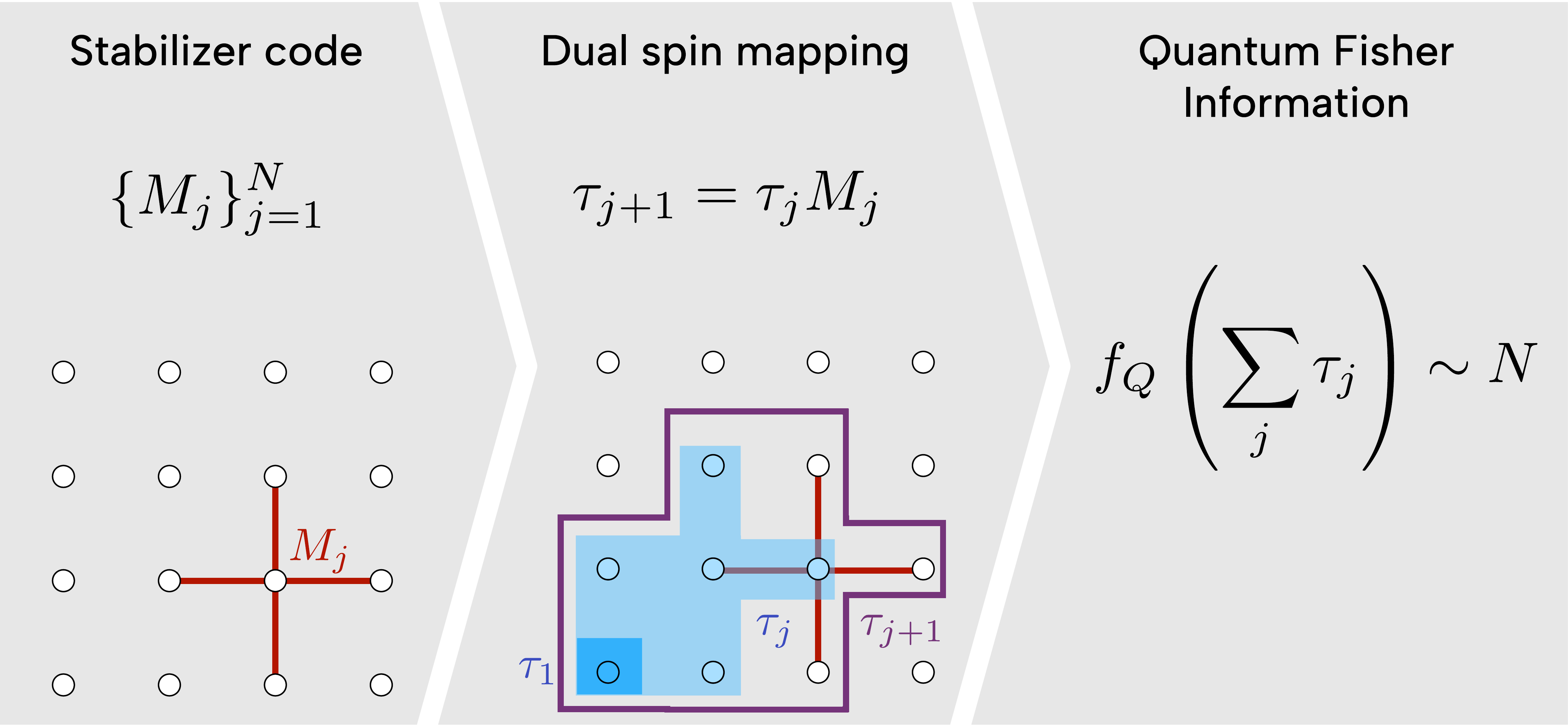}
 \caption{Schematic of the framework to find extensive QFI in stabilizers. \textit{Left}: A stabilizer code defined by a set of generators $\{M_j\}_{j=1}^{N}$ acting locally on a qubit lattice (illustrated by a 2D cluster states; red bonds denote the support of an illustrative $M_j)$. \textit{Center}: Construction of the dual spin operators $\tau_j$ via the recursive relation $\tau_{j+1} = \tau_j M_j$, starting from a unitary and hermitian operator $\tau_1$ which satisfies $[\tau_1, M_j]=0,\, \forall j$. Each resulting $\tau_j$ is a nonlocal, string-like operator whose support (blue shaded region) grows by absorbing successive stabilizers. \textit{Right}: The collective operator $\sum_j \tau_j$ yields an extensive QFI density $f_Q \sim N$.}
    \label{fig:intro}
\end{figure}

A natural route through these questions is offered by the quantum Fisher information (QFI) \cite{braunstein1997statistical, paris2009quantum, giovannetti2011advances, pezze2018quantum}, defined for a pure state $|\psi\rangle$ and an observable $ O$ as
$\mathcal{F}_Q( O) = 4\bigl[ \langle  O ^2 \rangle - \langle  O \rangle ^2\bigr]$.
This quantity has a clear operational meaning in metrology: the precision in estimating a parameter $\lambda$ imprinted by the unitary $U=e^{i\lambda  O}$ is bounded by the Cram\'er-Rao inequality, $(\Delta\lambda)^2 \geq 1/\mathcal{F_Q}$. In quantum information, the QFI becomes a witness of multipartite entanglement when $ O$ is a collective operator built from locally supported observables \cite{pezze2009entanglement, toth2014quantum, pezze2014quantumtheoryphaseestimation}. In particular, if the QFI density over $K$ qubits satisfies
\begin{equation}
f_\mathcal{Q} ( O) := \mathcal{F_Q}/K > m
\end{equation}
for $m < K$ integer, then at least $m+1$ parties are mutually entangled \cite{hylluys2012fisher, toth2012multipartite}. Recent work has used this metric to probe multipartite entanglement in monitored random Clifford circuits, finding that it is entirely absent unless a global symmetry provides a protection mechanism, even at criticality \cite{Solanilla2025multipartite}. However, the general conditions under which $f_\mathcal{Q}$ becomes extensive in systems with built-in symmetry or topological structure remain poorly understood, despite some studies on some specific Hamiltonians' ground states \cite{pezze2017multipartite, zhang2018characterization, zhang2022multipartite, ferro2025kicking,Zeng2015gappequantumliquid,Schuch2011projentpair}. This raises the question of what class of observables is needed to reveal extensive QFI in stabilizer codes.

In this work, we introduce a systematic framework to construct nonlocal observables that yield extensive QFI density across a broad class of stabilizer codes (cf.~Fig.~\ref{fig:intro} for a summary). Our construction maps stabilizer generators to dual Ising spins via a recursive relation, and shows that the two-point correlators of the dual variables are equivalent to string order parameters in the original microscopic description. This identification reveals that \textit{extensive QFI from the dual-spin observable is a metrological manifestation of long-range string order}~\cite{Affleck1987criticaltheory,Haldane1983continumdynamics,Haldane1983largespin,Pollmann2012SPT,Zeng2015gappequantumliquid,Schuch2011projentpair}. We showcase this framework on three paradigmatic stabilizer models, encompassing both SPT and intrinsically topological phases: (i) the one-dimensional cluster (XZX) code, (ii) the two-dimensional cluster code, (iii) the toric code. 
In the monitored setting, where stabilizer and single-site measurements compete, the QFI density undergoes transitions from an extensive to an intensive regime.

\section{General construction of non-local operators.}
Throughout this work, we consider a system of $K$ qubits, and we denote $L$ the linear dimension of the system.
The maximal scaling of the QFI density, $f_\mathcal{Q}=K$, is achieved by the Greenberger-Horne-Zeilinger (GHZ) state (and its locally rotated equivalents). 
The GHZ state is the ground state of a 1D Ising Hamiltonian  $H = \sum_{i=1}^{K-1} Z_i Z_{i+1}$, for which $(K=L)$.
In the stabilizer setting, these conditions are reproduced through stabilizer measurements $M_i = Z_i Z_{i+1}$ in a circuit. The extensive QFI arises from the observable $ O = \sum_i Z_i$, whose connected correlations satisfy $\langle Z_j Z_k\rangle =1$ and $\langle Z_j \rangle = 0$ for all $j,k$. Such finite connected correlations at arbitrary distances motivate the following protocol: given a stabilizer state $|\psi\rangle$ with stabilizers $\{ M_j \}_{j=1}^{N}$~\footnote{This statement corresponds to $M_j|\psi\rangle=|\psi\rangle$ for each $j=1,\dots,N$.}, we shall define dual spins $\tau_j$ such that:
\begin{equation}
    \left\{M_j = \tau_j \tau_{j+1} \; | \;\langle \tau_j \tau_k\rangle =1 \; \forall k,\;\langle \tau_j \rangle = 0\right\}_{j=1}^N,
\end{equation} 
inherits the GHZ-like correlation structure. 
This yields $f_Q \sim N$ for the observable $ O = \sum_j \tau_j$, which is extensive given $N\sim \mathcal{O}(K)$. 
Crucially, the same reasoning applies in monitored circuits, in which case $ O$ becomes a locally rotated version of the dual spins.

We establish a mapping between a specific class of stabilizer codes, comprising terms that commute with one another, and a one-dimensional Ising chain, leveraging the duality between quantum error-correcting codes and statistical mechanical models. In this framework, each site $j$ in the Ising chain corresponds to a dual Ising spin $\tau_j$, and these spins can be constructed utilizing the recursive relation:
\begin{equation}
\label{mapping}
    \tau_{j+1} = \tau_j M_j
\end{equation}
where $M_j$ represents the $j$-th term in the stabilizer code. This relation implies that $\tau_j$ is determined by the preceding spin $\tau_{j-1}$ and the term $M_{j-1}$ of the stabilizer code. 

Crucially, the above mapping 
should not define an additional
independent $\mathbb Z_2$ symmetry, but instead expose only the symmetry already
implicit in the original stabilizer code. Therefore, one must choose the initial dual spin $\tau_1$ appropriately and then follow the recursive formula to obtain the remaining dual Ising spins. The following properties are required: 
\begin{equation}
    \tau_1=\tau_1^\dagger=\tau_1^{-1} \hspace{10mm} \left[ \tau_1, M_j \right]=0,\;\forall j.
\end{equation}
They ensure that each $\tau_j$ is a Hermitian operator, preserving the observability of the system. Additionally, the $\tau_j$ operators at different sites commute with each other, and the relation
$
\tau_{j+1}^\dagger = M_j^\dagger \tau_j^\dagger = \tau_j M_j = \tau_{j+1}
$
holds if all the $M_j$'s in the stabilizer code commute with one another. Albeit non-necessary, it is often easy to choose $\tau_1$ as a local operator.
 We employ open boundary conditions for the rest of the analysis, unless explicitly stated. 
 
The QFI built from these nonlocal observables on the associated stabilizer states is given by the cardinality of the set of dual spins:
\begin{equation}
\label{qfi_cardi}
    \mathcal{F_Q} \left(  O=\sum_j \tau_j \right)=|\{\tau_j\}|^2.
\end{equation}
This construction encompasses the results for the QFI of ground states in Hamiltonian systems $H=\sum_j M_i$ discussed in Refs.~\cite{pezze2017multipartite, zhang2018characterization, zhang2022multipartite, ferro2025kicking}. \\

In monitored dynamics, measurement outcomes are stochastic: not all stabilizer measurements project the state onto the $+1$ eigenspace. The optimal observable therefore acquires trajectory-dependent signs, $ O (\mathbf n )= \sum_j n_j \tau_j$ with $n_j\in\{-1,1\}$. For a given trajectory $k$, we compute the QFI density $f_Q^{(k)}$ by optimizing over $\mathbf n=(n_1, \, n_2,\, \dots,\, n_N )$ using a classical simulated annealing algorithm \cite{paviglianiti2023multipartite, Solanilla2025multipartite}:
\begin{equation}
    f_Q^{(k)} = \max_{\mathbf n} f_Q^{(k)}\left( O \left(\mathbf n \right) \right)
\end{equation}
In what follows, stabilizer circuits are simulated in time polynomial in $L$ using the Gottesman-Knill framework \cite{gottesman1997stabilizer}. For each choice of parameters $L$ (system size) and $p$ (single-site measurement rate), we take at least $5\times10^3$ realizations of quantum trajectories to compute the QFI density of the stationary state. The annealing schedule converges quickly at low temperatures $T\sim 0.1$ over $\mathcal{O}(N^{3/2})$ iterations per temperature, with a constant step size flipping $n_j\to - n_j$. The typical QFI density is obtained by averaging over realizations: $f_Q \equiv \mathbb{E}_{k}[f_{Q}^{(k)}]$.

\section{Quantum Fisher information and long-range order}
The physical meaning of the dual-spin construction becomes transparent at the level of correlation functions. Starting from the defining relation $M_j=\tau_j\tau_{j+1}$,
one immediately obtains, for any interval $[a,b)$,
\begin{equation}
\prod_{j=a}^{b-1} M_j
=
\prod_{j=a}^{b-1} (\tau_j\tau_{j+1})
=
\tau_a\tau_b,
\end{equation}
since all intermediate dual spins cancel pairwise. This identity shows that a two-point correlator of the dual variables is exactly equivalent to a non-local string correlator~\cite{Affleck1987criticaltheory,Haldane1983continumdynamics,Haldane1983largespin,Pollmann2012SPT,Zeng2015gappequantumliquid,Schuch2011projentpair} in the original microscopic description:
\begin{equation}
\langle \tau_a\tau_b\rangle
=
\left\langle \prod_{j=a}^{b-1} M_j \right\rangle.
\end{equation}
Therefore, long-range order in the dual spins is nothing but long-range string order written in a dual language ~\cite{Affleck1987criticaltheory,Haldane1983continumdynamics,Haldane1983largespin,Pollmann2012SPT,Zeng2015gappequantumliquid,Schuch2011projentpair}.

This relation also clarifies why the quantum Fisher information constructed from the dual spins is sensitive to such hidden order. For the collective observable $
 O=\sum_{j=1}^{N}\tau_j,
$
the QFI of a pure state is
\begin{equation}
\mathcal F_Q( O)
=
4\,\mathrm{Var}( O)
=
4\sum_{j,k}
\Bigl(
\langle \tau_j\tau_k\rangle
-
\langle \tau_j\rangle\langle \tau_k\rangle
\Bigr).
\end{equation}
Whenever $\langle \tau_j\rangle=0$, this reduces to
\begin{equation}
\mathcal F_Q( O)
=
4\sum_{j,k}\langle \tau_j\tau_k\rangle
=
4\sum_{j,k}
\left\langle
\prod_{\ell=j}^{k-1} M_\ell
\right\rangle.
\end{equation}
In this sense, the dual-spin construction transforms hidden non-local order into a form that can be directly accessed through a collective observable due to the emergence of the GHZ state in terms of dual variables.

The physical interpretation is therefore straightforward: if the string correlator decays rapidly with distance, only nearby pairs contribute, and the QFI remains at most extensive ($f_Q\sim O(1)$). By contrast, if the string correlator approaches a finite constant at long distances, then a macroscopic number of pairs contributes coherently, leading to a parametrically enhanced QFI ($f_Q\sim O(N)$). The large QFI extracted from the dual-spin observable should therefore be understood as a metrological manifestation of long-range string order. 

\section{1D Monitored cluster code.}
We start our analysis by investigating one-dimensional monitored cluster codes \cite{raussendorf2001oneway, lang2020entanglement} exhibiting a transition from a long-range multipartite entangled phase to an area law phase in the $\tau$-structure. 
More concretely, the system is initialized in the state $\lvert 0 \rangle^{\otimes L}$, with $ L $ even. Recall, in this case, $K=L=N+2$. At each update step, we perform either a stabilizer measurement
\begin{equation}
    M_j = X_{j-1} Z_j X_{j+1}
\end{equation} with probability $ 1 - p_z $ for a random site $ j \in \{2, \dots, L-1\} $, or a single-body $ Z_j $ measurement with probability $ p_z $ for a random $ j \in \{1, \dots, L\} $. The full evolution consists of $ 2L $ time steps, each comprising $ L $ update steps.

Following the recursive construction, one can choose $\tau_1 = -Z_1X_2$ as the initial dual spin \cite{ferro2025kicking}, satisfying $\tau_1 = \tau_1^\dagger = \tau_1^{-1}$ and $[\tau_1, M_j]=0$ for all $j$. Recursively, we obtain
\begin{equation}
\begin{split}
    \tau_2 &= \tau_1 M_1 = -Y_1 Y_2 X_3 \\
    \tau_3 &= \tau_2 M_2 = Y_1 Z_2 Y_3 X_4\\
    \vdots \\
    \tau_j &= -(-1)^j Y_1 \prod_{n=2}^{j-1}Z_nY_jX_{j+1} \hspace{5mm} \text{for }1<j<L.
\end{split}
\end{equation}
The operators are string-like, with support growing linearly in $j$.

\begin{figure}
    \centering
    \includegraphics[width=\linewidth]{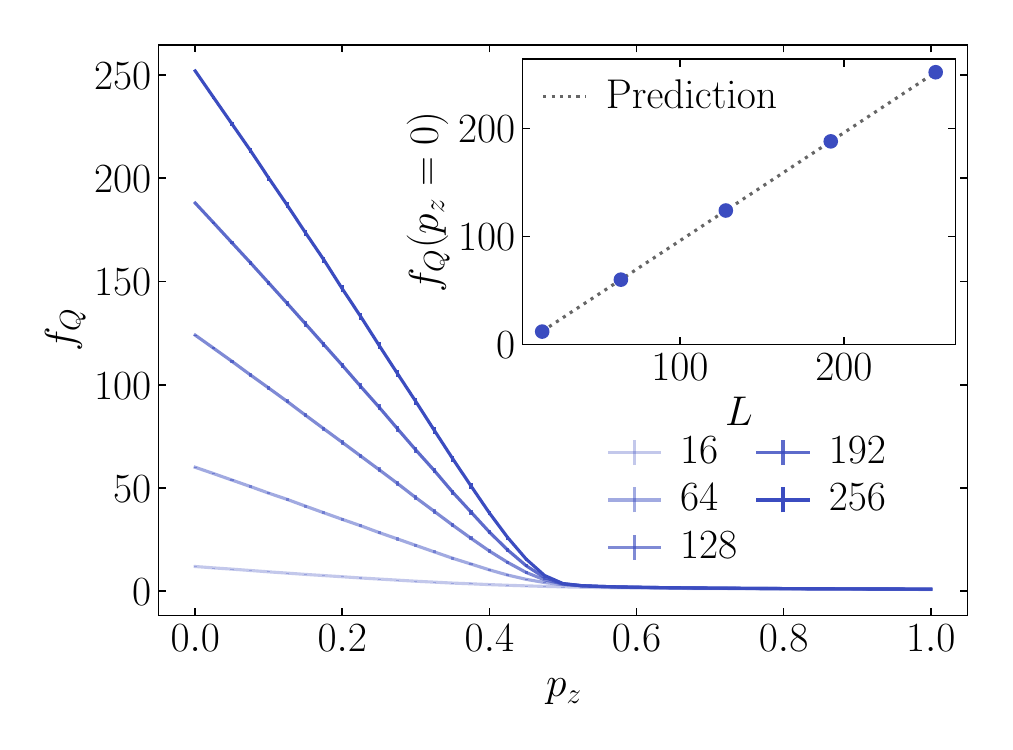}
    \vspace{-10mm}
    \caption{QFI density of the monitored one-dimensional cluster code versus local measurement rate $p_z$ for various $L$'s, showing a change in scaling from extensive to intensive at $p_z\sim 0.5$. (Inset) The scaling of the QFI density at $p_z=0$ and its theoretical match given by the cardinality of the set of dual spins. }
    \label{fig:1D_cluster_qfi_pz}
\end{figure}
In Fig. \ref{fig:1D_cluster_qfi_pz}, we show how the QFI density $f_Q$ changes with the measurement rate $p_z$. We find extensive QFI density, $f_Q\sim \mathcal O(L)$, at low $p_z$ rates. This is the expected result when stabilizer measurements are dominant. Instead, at high single-body measurement rates, $Z_j$ measurements dominate the dynamics, and $f_Q$ becomes intensive: $f_Q \sim \mathcal O(1)$.

Studying in more detail the case of $p_z=0$, in the inset of Fig. \ref{fig:1D_cluster_qfi_pz}, we compare the numerical scaling of $f_Q$ with the theoretical prediction (dashed line). 
The extensive scaling follows from the dual–spin mapping: in the $\tau$-language the cluster-state generators reduce to nearest-neighbor terms $\tau_j\tau_{j+1}$, so the collective operator $O=\sum_{j} \tau_j$ yields quadratic QFI with respect to the cardinality of the set $\{\tau_j\}$ for $1<j<L-2$: $\mathcal{F_Q}\sim |\{\tau_j\}|^2$.
Hence, the predicted QFI density satisfies Eq.\eqref{qfi_cardi} and reads
\begin{equation}
    f_Q (p_z=0) = \frac{\left|\{\tau_j\}_{j=1}^{L-2}\right|^2}{L} = \frac{(L-2)^2}{L}.
\end{equation}

For large $p_z$, i.e., in the regime dominated by single-site measurements, the QFI remains finite: being a sum of connected correlators of the dual $\tau$ variables, which decay exponentially in the presence of a finite, measurement-induced correlation length; consequently, the QFI density stays finite in this parameter range. 

Remarkably, we observe a transition from extensive QFI density, $f_Q \propto L$, to an intensive phase at $p_z = 0.5$ \cite{lavasani2021measurement, sang2021measurementprotected, klocke2022topological}, as corroborated by the scaling analysis inferred from the topological entanglement in the End Matter (cf.~Fig.~\ref{fig:1Dcluster_EndMat}). 
Notably, at criticality, we do not find extensive scaling of the QFI. This can be argued with the help of scale-invariant two-point correlators of local observables, $C_{i,j}\sim |i-j|^{-2\Delta}$, for which the QFI density diverges with system size only if $\Delta<\tfrac{1}{2}$ in one dimension \cite{hauke2016measuring, Solanilla2025multipartite, zabalo2020critical, zabalo2022operator}; hence a finite QFI at $p_z\approx 0.5$ can in principle arise if the effective exponent is $\Delta>\tfrac{1}{2}$ for the connected correlators of the dual spins.

\section{2D Monitored cluster code.}

We extend our analysis to two-dimensional (2D) monitored cluster codes defined on a square lattice of side length $L$, with $N=L^2$ qubits. The system is initialized in the state $\lvert \psi \rangle = \lvert 0 \rangle^{\otimes L^2} $. At each update step, we perform one of the following measurements: with probability $ p_y $, we perform a one-body $ Y_{i,j} $ measurement on a randomly chosen site $(i,j) $ with $ i,j \in \{1, \dots, L\} $; or with probability $1-p_y$, we measure the stabilizer operator
\begin{equation}
M_{i,j} = X_{i,j} \prod_{\langle i',j' \rangle \sim \langle i,j \rangle} Z_{i',j'},
\end{equation}
where $ X_{i,j} $ acts on a randomly chosen site $(i,j)$ with $i,j \in \{2, \dots, L-1\} $, and the product runs over all nearest neighbors $\langle i',j' \rangle $ of $(i,j) $, with $ Z_{i',j'} $ acting on each nearest neighbor. A single time step consists of $ L^2 $ consecutive update steps, and the system is evolved for $t=2L^2$ time steps. The same dual-spin analysis reveals a transition between a long-range entangled phase and an area law phase, characterized by the scaling of $f_Q$ built from the nonlocal observables.

\begin{figure}[htb]
    \centering
    \includegraphics[width=\linewidth]{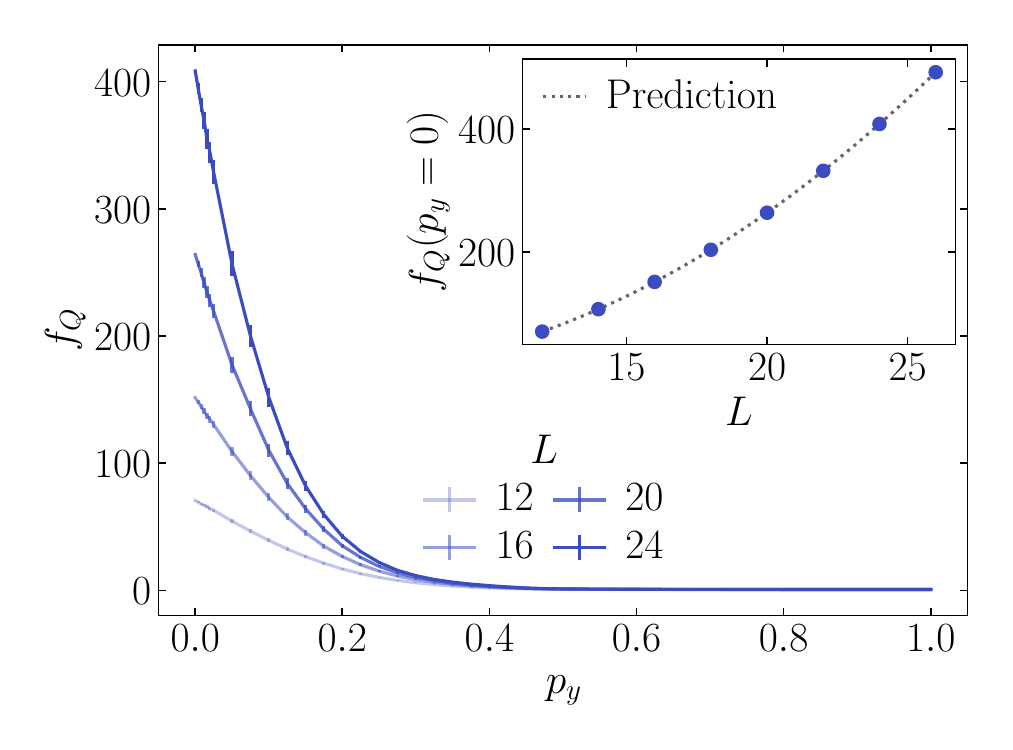}
    \vspace{-10mm}
    \caption{QFI density of the monitored two-dimensional cluster code versus the single-site measurement rate $p_y$ for various $L$, revealing a transition from extensive to intensive scaling near $p_y \simeq 0.5$. (Inset) Scaling of the QFI density at $p_y=0$ and its theoretical prediction, showing quadratic growth with $L$.}
    \label{fig:2D_cluster_qfi_py}
\end{figure}
\begin{figure}[h!]
\centering
\includegraphics[width=\linewidth]{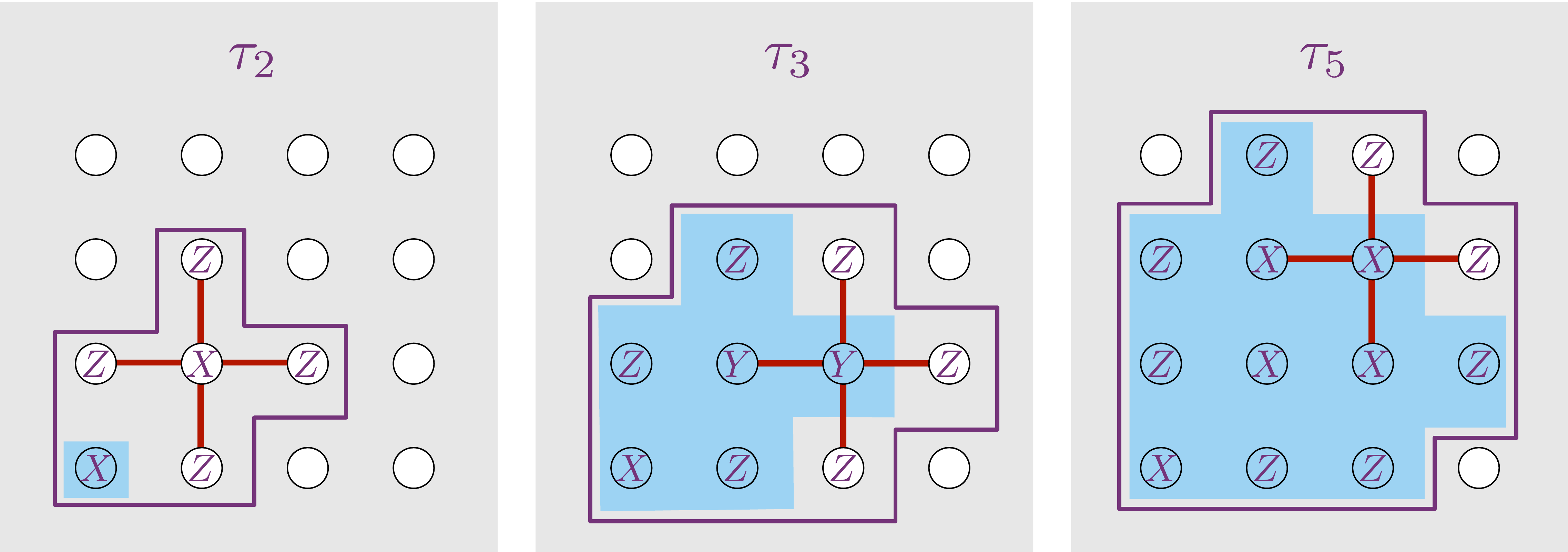}
\caption{Examples of the dual spin operators used in a square lattice cluster code of $L=4$, $\tau_2$, $\tau_3$, and $\tau_5$. The support of $M_j$, $\tau_j$, and $\tau_{j+1}$ is indicated by red links, blue shadow, and purple contour, respectively.}
\label{fig:2D_cluster_observables}
\end{figure}

As shown in Fig. \ref{fig:2D_cluster_qfi_py}, the emergence of $f_Q \sim L^2$ scaling at $p_y = 0$ and area law phase for high $p_y$ can again be explained in terms of dual spin operators, as discussed in the 1D case. Here, in a 2D square lattice with open boundary conditions, we have chosen $\tau_1=X_{1,1}$ to lie at the lower-left lattice corner. Given that the stabilizers now are star-shaped, none of them have support in the corner qubits, and therefore the initialization of all the qubits in the system as $|0\rangle^{\otimes L^2}$ ensures that $\langle\tau_1\rangle=0$ throughout the dynamics. The recursive formula yields $\tau_2=\tau_1 M_{2,2}$, $\tau_3 = \tau_2 M_{3,2}$, and so on until $\tau_{(L-2)^2+1}$. In Fig.~\ref{fig:2D_cluster_observables} we show representative examples of these operators.

In this case, we have $(L-2)^2$ stabilizers $M_{i,j}$ and the predicted QFI density is
\begin{equation}
    f_Q (p_y=0) = \frac{\left|\{ \tau_j\}_{j=1}^{(L-2)^2+1}\right|^2}{L^2} = \frac{(L^2 - 4L + 5)^2}{L^2} \ .
\end{equation}
The QFI density captures the measurement-induced phase transition identified by bipartite entanglement (see Fig.~\ref{fig:2D_cluster_tmi} in the End Matter) through the extensive and intensive scalings at low and high $p_y$, respectively. 

\section{Toric Code.}
Finally, we extend our study to a two-dimensional system with intrinsic topological order, the toric code \cite{kitaev2003fault, lavasani2021topological}, on an $L\times L$ square lattice with spins on edges. We study a stochastic measurement–update scheme applied for $t = 2 L^{2}$ steps to reach a stationary state, where each time step consists of $2L^2$ operations. Starting from $\ket{0}^{\otimes 2L^2}$, each update either measures a four-body stabilizer $M_{i,j}$ (with probability $1-p_{y}$), where the stabilizer is either a star operator $A_s$ acting on the four edges incident to a uniformly random interior vertex, or a plaquette operator $B_p$,
\begin{equation}
    A_{s}=\prod_{(i,j)\in s} X_{(i,j)},
\hspace{1cm}
    B_{p}=\prod_{(i,j)\in p} Z_{(i,j)}.
\end{equation}
Otherwise, a single-spin Pauli-$Y$ measurement $Y_{i,j}$ is performed on a uniformly random spin (with probability $p_{y}$).

\begin{figure}[h!]
\centering
\includegraphics[width=\linewidth]{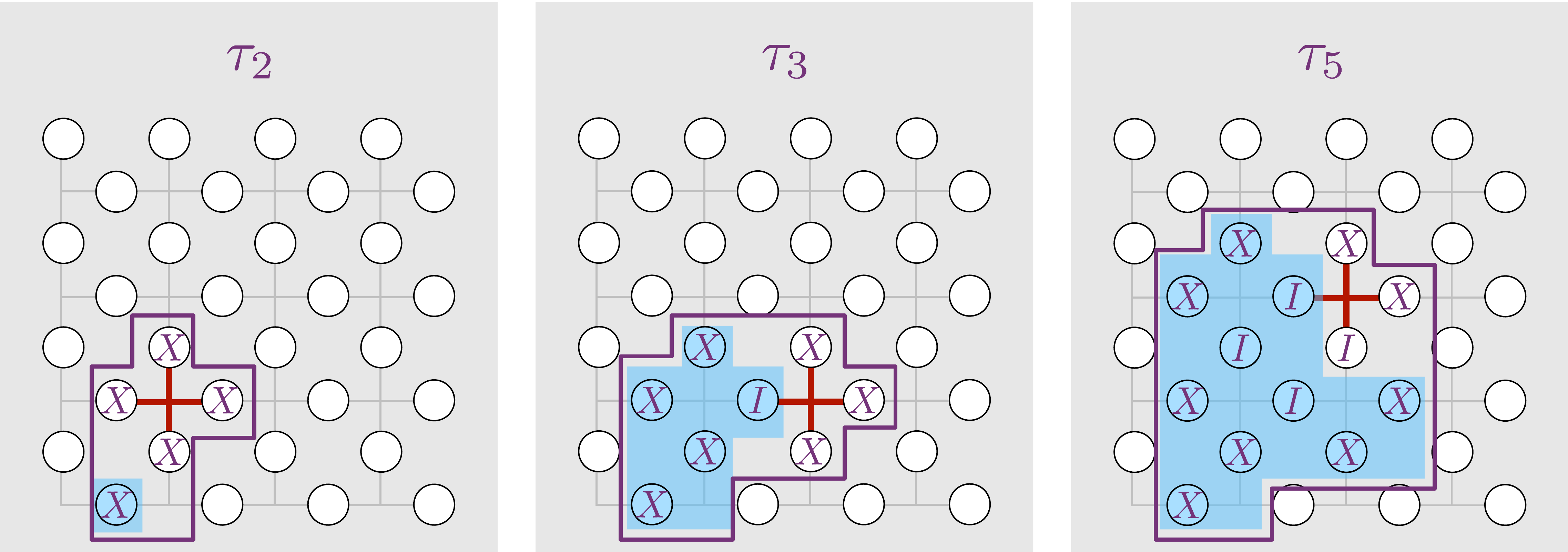}
\caption{Examples of $\tau_2$, $\tau_3$, and $\tau_{5}$ in a square lattice toric code with $L=4$. The support of $M_j$, $\tau_j$, and $\tau_{j+1}$ is indicated by red links, blue shadow, and purple contour, respectively.}
\label{fig:tc_observables}
\end{figure}

In this case, the dual spins are chosen from a subset of all the $A_s$ operators~\footnote{Such a subset consists of all star operators $A_s$ with no support on boundary spins. An equivalent picture can be done using only plaquette operators $B_p$, which shows the potential of this protocol despite this freedom of choice.}. See Fig.\ref{fig:tc_observables} for some examples and the End Matter for the construction. We have $(L-2)^2$ stabilizers $M_{i,j}$ and the predicted QFI density is
\begin{equation}
    f_Q (p_y=0) = \frac{\left|\{ \tau_j\}_{j=1}^{(L-2)^2+1}\right|^2}{2L^2} = \frac{(L^2 - 4L + 5)^2}{2L^2}\ .
\end{equation}

\begin{figure}[htb]
    \centering
    \includegraphics[width=\linewidth]{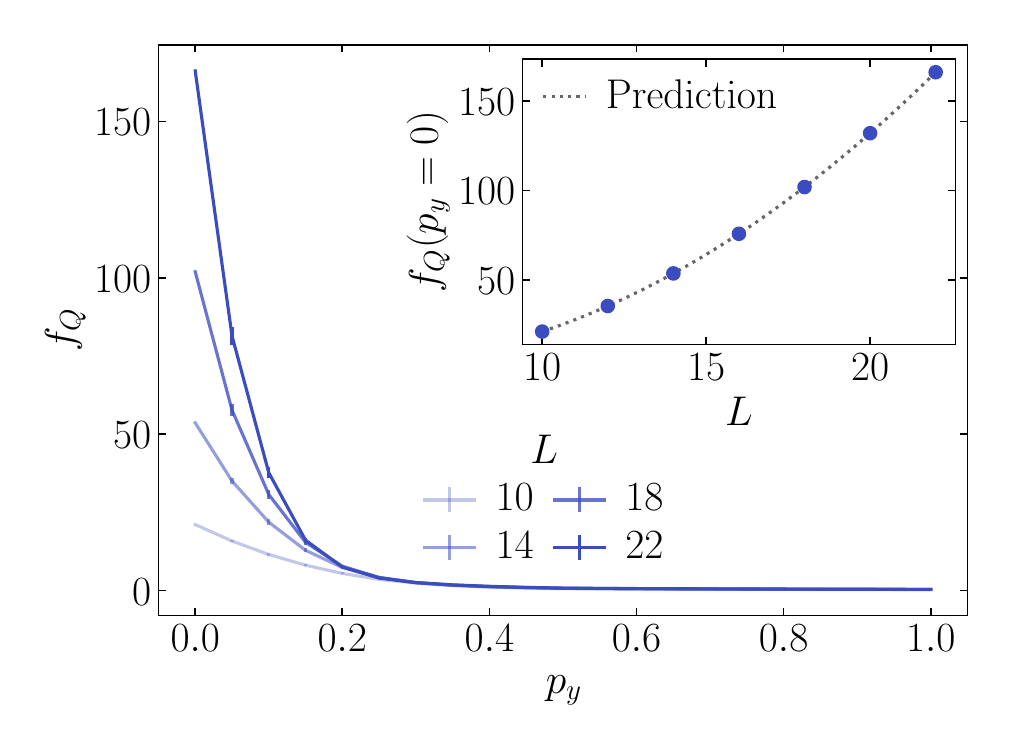}
    \vspace{-10mm}
    \caption{QFI density of the monitored toric code with $A_s$ stabilizers and single-qubit $Y$ measurements, for various $L$, revealing a transition from extensive to intensive $f_\mathcal{Q}$ as $p_y$ is increased. (Inset) Scaling of the QFI density at $p_y=0$ and its theoretical prediction, showing quadratic growth with $L$.}
    \label{fig:tc_qfi_py}
\end{figure}

In this setting, the QFI density $f_{Q}$ again exhibits a transition from an extensive ($f_Q \propto L^2$) to an intensive phase, as shown in Fig.~\ref{fig:tc_qfi_py}. For the toric code, our dual mapping results in a 2D Ising model, yielding QFI density scaling with the $\sim L^2$ sites, unlike
constructions giving rise to $L$ decoupled Ising chains, which give only $L$
scaling~\cite{zhang2018characterization}.

\section{Discussion and Conclusions.}
We introduced a systematic framework to construct operators that yield extensive quantum Fisher information density across a broad class of stabilizer codes, encompassing both symmetry-protected topological phases and intrinsically topological orders \cite{paviglianiti2023multipartite, poggi2024measurementinduced, sharma2025multipartite}. 
We apply this framework to monitored settings: in the 1D cluster model, the QFI probes the underlying transition. Instead, in 2D, finite-size effects become too severe.

Crucially, one of the central findings of this work is that the extensive scaling of the QFI associated with the dual-spin observable 
$O=\sum_j \tau_j$ reflects the connection 
between quantum Fisher information and nonlocal long-range correlations, as made explicit 
by our construction.

At the same time, our results indicate that multipartite entanglement in stabilizer systems cannot be directly 
inferred from QFI measures alone. In particular, the QFI constructed 
from local operators (see End Matter) exhibits only intensive scaling, consistent with 
the conclusions of Ref.~\cite{ferro2025kicking}. This observation highlights the importance 
of combining nonlocal QFI-based approaches \cite{zhang2018characterization, 
zhang2022multipartite} with complementary entanglement probes 
\cite{pezze2017multipartite, frerot2018quantum, costa2021from, vitale2023robust, 
laurell2021quantifying, scheie2021witnessing}.

Several avenues remain open. Exploiting exactly solvable mappings in monitored code architectures could yield analytic control of the QFI at critical points \cite{lang2020entanglement, klocke2023majorana}. Extending the framework to include feedback, weak, or continuous measurements \cite{passarelli2024manybody, li2025monitoredlongrangeinteractingsystems} would test whether QFI-defined phase boundaries coincide with purification and decoding thresholds. Finally, implementing these nonlocal generators into concrete metrology protocols on current platforms \cite{yu2022quantum, rams2018at, yaoming2021dynamic} constitutes a promising direction for bridging the gap between theoretical diagnostics and experimental quantum advantage.

\medskip 

\begin{acknowledgments} 
\textit{Acknowledgments.} We thank X. Chen and S. Trebst for useful discussions. We are especially thankful to B. Han for insightful comments and conversations. We acknowledge funding by the Deutsche Forschungsgemeinschaft (DFG, German Research Foundation) under Projektnummer 277101999 - TRR 183 (project B01 and B02), and under Germany's Excellence Strategy - Cluster of Excellence Matter and Light for Quantum Computing (ML4Q) EXC 2004/1 - 390534769. The most demanding numerical simulations were performed on the RAMSES cluster at RRZK Cologne.

\textit{Data Availability.} The code of our data is available at Ref.~\cite{dataavail}. 
\end{acknowledgments}

\bibliography{biblio}
\bibliographystyle{apsrev4-2}

\clearpage

\setcounter{section}{0}
\setcounter{secnumdepth}{2}

\begin{widetext}
\begin{center}
\textbf{End matter}
\end{center}

\section{Scaling analysis of QFI density}

To quantify the transitions in QFI scaling reported in the main text, we extract scaling exponents by fitting $f_Q$ as a function of system size at each measurement rate. The results for the 1D cluster model are summarized in Fig.~\ref{fig:1D_fit}.

In the 1D cluster code, the scaling exponents confirm a sharp transition from extensive to intensive QFI at $p_z^c\approx 0.5$, consistent with the bipartite entanglement transition (Fig.~\ref{fig:1D_cluster_stopo}). In the 2D cluster code and the toric code, the QFI scaling transition could not be extracted by fitting the curve $a+bL^c$ due to finite-size effects. A rough comparison of Figs.~\ref{fig:2D_cluster_qfi_py} and \ref{fig:tc_qfi_py} with \ref{fig:2D_cluster_tmi} and \ref{fig:tc_tmi} seems to indicate that, in this case, the transition point between QFI density phases and the one found via bipartite entanglement does not coincide.

\medskip

\section{QFI in terms of local observables}

Here we show that the QFI density built from local (single-site) observables remains finite across all phases. Following Ref.~\cite{Solanilla2025multipartite}, we define $F_Q \equiv \max_{\{\mathbf n_j\}} F_Q\!\bigl( O^{\{\mathbf n\}}\bigr)$ with $ O^{\{\mathbf n\}} = \sum_j \mathbf{n}_j \cdot \boldsymbol{\sigma}_j$, and optimize over local directions $\{\mathbf{n}_j\}$ using a classical annealing routine. For each pair $(L,p_z)$, we average over at least $5\times10^3$ quantum trajectories. As shown in Fig.~\ref{QFI_local}, the QFI remains finite across both phases and at the critical point, confirming that strictly local probes cannot diagnose this class of monitored stabilizer codes.

\begin{figure}[h]
    \centering
    \begin{subfigure}{0.4\linewidth}
        \phantomcaption
        \stackinset{l}{5pt}{t}{10pt}{(\thesubfigure)}{
        \includegraphics[width=\textwidth]{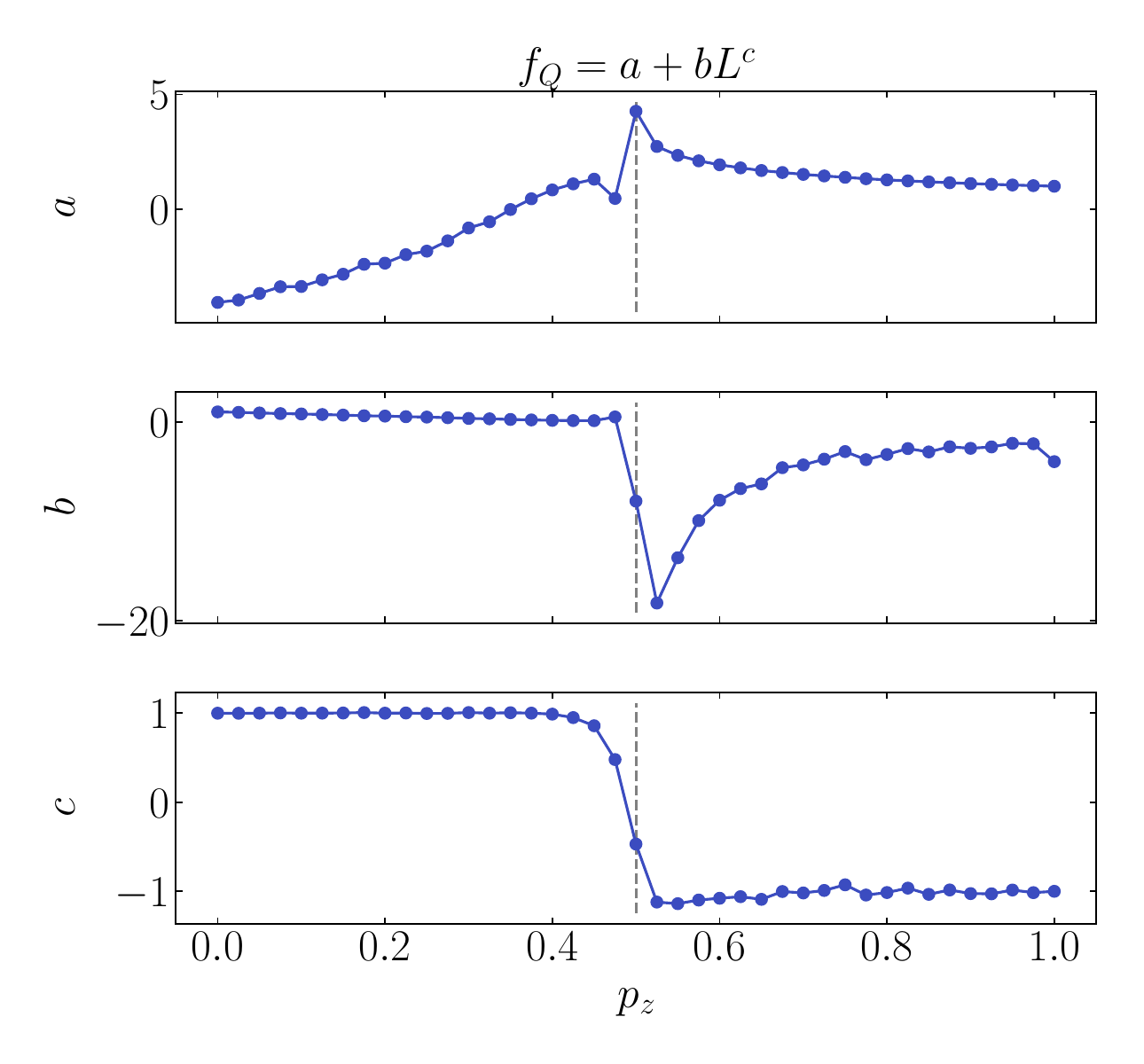}
        }
        \label{fig:1D_fit}
    \end{subfigure}
    \hspace{5mm}
    \begin{subfigure}{0.4\linewidth}
        \phantomcaption
        \stackinset{l}{5pt}{t}{10pt}{(\thesubfigure)}{\includegraphics[width=\textwidth]{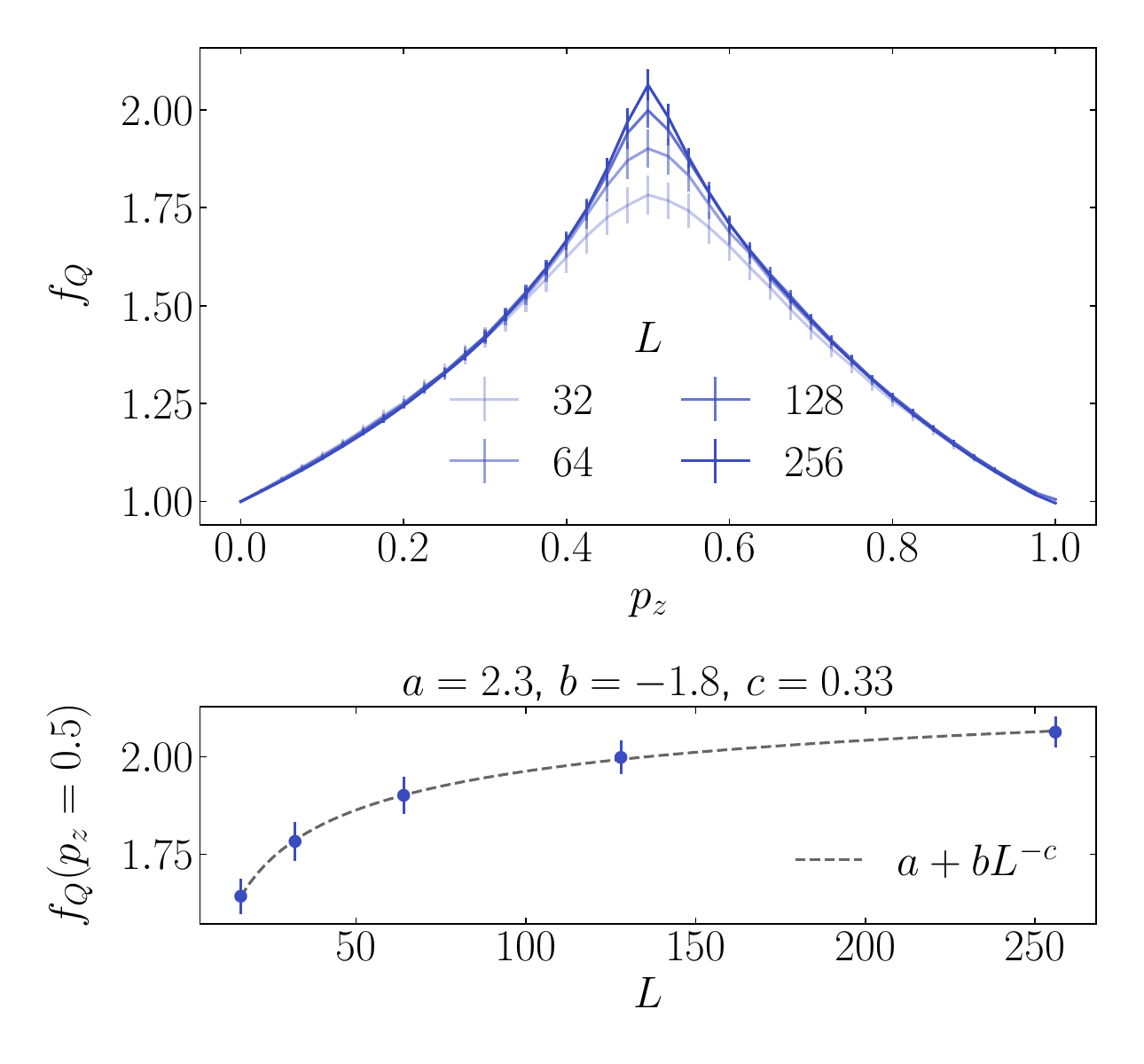}
        }
        \label{QFI_local}
    \end{subfigure}
\vspace{-5mm}
\caption{Additional data of the QFI density in the 1D cluster code. (a)~Scaling exponent extracted from the QFI density for the 1D cluster code, revealing a transition at $p_z = 0.50$ (dashed line). (b)~QFI density of local collective operators in a 1D measurement-only cluster code with single-qubit $z$ measurements, for various system sizes $L$. Neither the SPT phase, the area law phase, nor the critical point can be differentiated by strictly local operators (top panel). The finiteness of the QFI is corroborated by its scaling with $L$ (bottom panel).}
\label{fig:1Dcluster_EndMat}
\end{figure}

\medskip

\section{Bipartite entanglement MIPTs}

We corroborate the QFI phase diagrams with independent bipartite entanglement measures for each model.

\underline{1D cluster state.}---We use the conditional mutual information $\mathcal{I}_2(A, C | B) = S_{A\cup B} + S_{B\cup C} - S_B - S_{A\cup B\cup C}$ on a partitioning of the system into $A, B, D, C$ with $C$ spatially separated from $A$ and $B$, where $S_K= -\operatorname{Tr}(\rho_K \log_{2}\rho_K)$ is the entanglement entropy. This quantity defines $S^{q}_{\mathrm{topo}}$ \cite{klocke2022topological}, which equals $2$ in SPT phases and $0$ in the trivial area law phase. As shown in Fig.~\ref{fig:1D_cluster_stopo}, we find the critical point at $p_z^c=0.5$ with $\nu=1.31(3)$, consistent with 2D classical percolation universality \cite{essam1980percolation, lang2020entanglement} ($\nu=4/3$).

\underline{2D cluster state.}---In Fig.~\ref{fig:2D_cluster_tmi}, we validate the phase diagram using the tripartite mutual information (TMI) \cite{zabalo2020critical}, $\mathcal{I}_{3}=S_A + S_B +S_C +S_{A\cup B \cup C}-S_{A\cup B}-S_{B\cup C}-S_{C\cup A}$. The critical point is found at $p_{y}\approx 0.51(2)$, and the QFI density at criticality remains intensive, consistent with the main text.

\underline{Toric code.}---The TMI as a function of the $Y$-measurement rate shows a corresponding transition near $p_{y}\approx 0.5$ (Fig.~\ref{fig:tc_tmi}).

\begin{figure}[h!]
    \centering
    \begin{subfigure}{0.30\linewidth}
        \phantomcaption
        \stackinset{l}{5pt}{t}{10pt}{(\thesubfigure)}{
        \includegraphics[width=\textwidth]{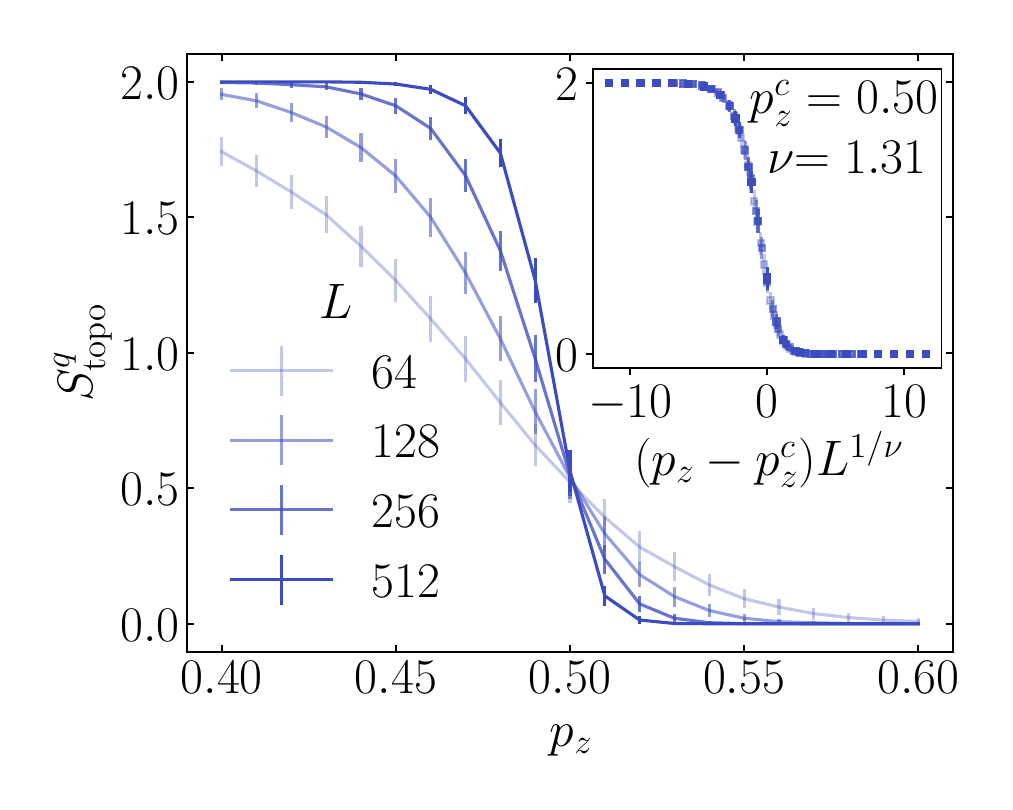}
        }
        \label{fig:1D_cluster_stopo}
    \end{subfigure}
    \hspace{3mm}
    \begin{subfigure}{0.30\linewidth}
        \phantomcaption
        \stackinset{l}{5pt}{t}{10pt}{(\thesubfigure)}{
        \includegraphics[width=\textwidth]{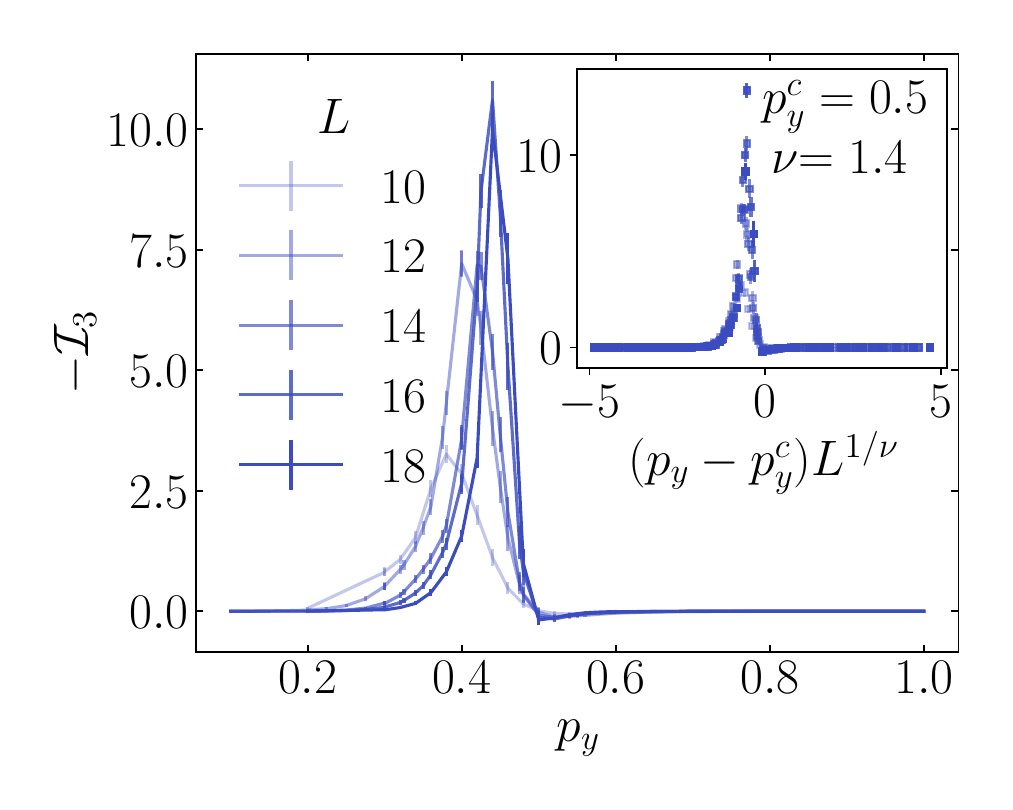}
        }
        \label{fig:2D_cluster_tmi}
    \end{subfigure}
    \hspace{3mm}
    \begin{subfigure}{0.30\linewidth}
        \phantomcaption
        \stackinset{l}{5pt}{t}{10pt}{(\thesubfigure)}{
        \includegraphics[width=\textwidth]{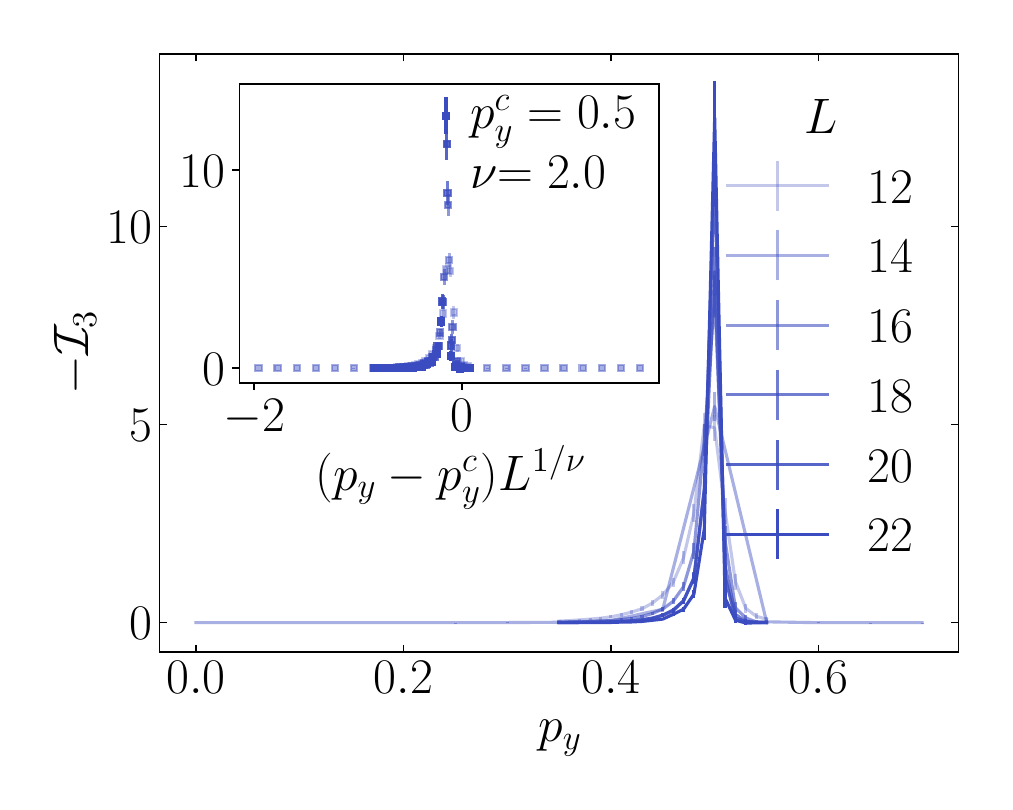}
        }
        \label{fig:tc_tmi}
    \end{subfigure}
\vspace{-5mm}
\caption{(a)~$S^{q}_{\mathrm{topo}}$ of the 1D cluster state around the transition point with finite-size scaling: $p_z^c = 0.5$, $\nu=1.31(3)$. (b)~Tripartite mutual information $\mathcal{I}_{3}$ for the 2D cluster state: $p_y^c=0.51(2)$, $\nu=1.4(5)$. (c)~$\mathcal{I}_{3}$ for the toric code: $p_y^c=0.5(1)$, $\nu=2(2)$.}
\label{bipartite entanglement}
\end{figure}

\medskip

\section{Dual spins construction in the toric code}

As we have shown graphically in Fig.\ref{fig:tc_observables}, the dual spins constructed with observables $A_s$ of the toric code are generated starting from $\tau_1 = X_{0,0}^h$. The subindices indicate the pairs \{row, column\}, and $h$ that it is a horizontal bond instead of a vertical one ($v$). Following the recursive relation introduced in the main text, the rest of the dual spins are obtained as follows:

\begin{equation}
    \tau_2 = \tau_1 A_s^{1,1} = X_{0,0}^h( X_{1,0}^h X_{1,1}^h X_{0,1}^v X_{1,1}^v)
\end{equation}
\begin{equation}
    \tau_3 = \tau_2 A_s^{1,2} = X_{0,0}^h( X_{1,0}^h X_{1,1}^h X_{0,1}^v X_{1,1}^v) (X_{1,1}^h X_{1,2}^h X_{0,2}^v X_{1,2}^v) = X_{0,0}^h X_{1,0}^h X_{0,1}^v X_{1,1}^v X_{1,2}^h X_{0,2}^v X_{1,2}^v,
\end{equation}
and so forth. Notice how, in the main text, we have illustrated and counted the dual spins in a way that no star operators $A_s$ on the boundaries have been used. This allows us to showcase the exact dependency of the QFI density saturation value on the cardinality of the set of dual spins generated.
\end{widetext}

\end{document}